# Engineering of anomalous Josephson effect in coherently coupled Josephson junctions

Sadashige Matsuo[1], Takaya Imoto[1,2], Tomohiro Yokoyama[3], Yosuke Sato[1], Tyler Lindemann[4], Sergei Gronin[4], Geoffrey C. Gardner[4], Michael J. Manfra[4,5,6,7], Seigo Tarucha[1,8]

[1] *Center for Emergent Matter Science, RIKEN, Saitama 351-0198, Japan*
[2] *Department of Physics, Tokyo University of Science, Tokyo 162-8601, Japan*
[3] *Department of Material Engineering Science, Graduate School of Engineering Science, Osaka University, Osaka 560-8531, Japan*
[4] *Birck Nanotechnology Center, Purdue University, West Lafayette, Indiana 47907, USA*
[5] *Department of Physics and Astronomy, Purdue University, West Lafayette, Indiana 47907, USA*
[6] *School of Materials Engineering, Purdue University, West Lafayette, Indiana 47907, USA*
[7] *Elmore Family School of Electrical and Computer Engineering, Purdue University, West Lafayette, Indiana 47907, USA*
[8] *RIKEN Center for Quantum Computing, RIKEN, Saitama 351-0198, Japan*

**A Josephson junction (JJ) is a key device in the development of superconducting circuits, wherein a supercurrent in the JJ is controlled by the phase difference between the two superconducting electrodes. Recently, it has been shown that the JJ current is nonlocally controlled by the phase difference of another nearby JJ via coherent coupling. Here, we use the nonlocal control to engineer the anomalous Josephson effect. We observe that a supercurrent is produced by the nonlocal phase control even without any local phase difference, using a quantum interference device. The nonlocal phase control simultaneously generates an offset of a local phase difference giving the JJ ground state. These results provide novel concepts for engineering superconducting devices such as phase batteries and dissipationless rectifiers.**

Symmetry-breaking superconducting (SC) junctions have often been used to explore novel SC phenomena including SC diodes [1–3], topological superconductivity, and Majorana zero modes [4]. The Josephson junction (JJ) [5] is an SC device that is often utilized for searching such SC phenomena. The JJs exhibit anomalous Josephson effect (AJE) upon the disorientation of the time-reversal and spatial-inversion symmetries, in which a finite phase difference away from 0 and $\pi$ produces the ground state of the JJ, called φ junctions [6–9]. Consequently, a spontaneous supercurrent emerges and the supercurrent is present even at zero phase difference. The φ junction has recently attracted considerable attention for manifold applications in SC phase batteries [10,11] and SC diodes [1,12–15]. To date, various systems have been proposed for realizing the φ junctions such as JJs of s-wave SC metals in the presence of spin-orbit interactions and magnetic fields [6–9], and have been experimentally verified as well [1,11,16,17]. However, strong magnetic fields or ferromagnetism are required to disintegrate the time-reversal symmetry in the proposed systems, but they can degrade the superconductivity.

Recently, it has been proposed that short-range coherent coupling of two JJs sharing one SC electrode can hybridize the Andreev bound states [18–21] in the respective JJs to form Andreev molecule states (AMSs). The coherent coupling through the shared SC electrode is intermediated by elastic cotunneling or crossed Andreev reflection [22–27], and the SC transport has been studied theoretically and experimentally [28–31]. The short-range coupling of two JJs generates the nonlocal Josephson effect, where the supercurrent in a JJ depends on the local phase difference as well as the nonlocal phase difference of the other JJ [28].

In the coupled JJs illustrated in Fig. 1(a), the supercurrent in the JJ1 satisfies the time-reversal relation of $I_{sc1}(\phi_1, \phi_2) = -I_{sc1}(-\phi_1, -\phi_2)$, where $\phi_1$ and $\phi_2$ denote the phase differences in JJ1 and JJ2, respectively. Upon setting $\phi_2$ and considering only JJ1, a character of the φ junction can emerge [28] because the time reversal and spatial inversion symmetries can be regarded as disintegrated at $\phi_2 \neq 0, \pi$. This realization method is qualitatively distinct from previous proposals because it requires neither strong magnetic fields nor ferromagnetic materials. Furthermore, the φ junction and spontaneous supercurrent of JJ1 obtained by such a mechanism are nonlocally controllable through JJ2, which will provide a new control method for phase batteries, or SC diodes. We have previously demonstrated the SC diode effect derived from the coherent coupling. However, the AJE (spontaneous supercurrent and φ junction) has not yet been experimentally verified.

In this study, we present an experimental demonstration of the spontaneous supercurrent and φ junction derived from the coherent coupling between two planar JJs. Planar JJs provide an experimental platform for topological superconductivity [32–34], and the coherent coupling between JJs allows us to engineer the topological superconductivity in two-dimensional systems [35,36]. Therefore, the phase-engineering of the AJE we demonstrated in the planar JJs develops our understanding of topological superconductivity and its phase engineering.

To demonstrate the phase engineering of the AJE, we need to evaluate the two-dimensional current phase relation (CPR) $I_{sc1}(\phi_1, \phi_2)$, i.e., the supercurrent in JJ1, as a function of the local and nonlocal phase differences. For this sake, we utilize asymmetric superconducting quantum interference devices (SQUIDs) [37] whose scanning electron microscope (SEM) image and schematic are shown in Figs. 1(b) and (c), respectively. The device is fabricated on a high-quality InAs quantum well covered with an epitaxial aluminum thin film. The stacking of the epitaxial aluminum and the InAs quantum well provides a highly transparent interface to serve an ideal platform for studying the physics of superconductor-semiconductor junctions [38–40]. The device includes the coupled JJs (JJ1 and JJ2) with a shared SC electrode of width 150 nm. The separation between JJ1 and JJ2 is sufficiently shorter than the coherence length of aluminum (~ 1 μm). We note that the Andreev molecule states have been demonstrated in coupled JJs with the same separation [41]. The junction length and width of the JJ1 and JJ2 are 100 and 600 nm, respectively. Furthermore, two larger JJs, named JJL1 and JJL2 with 2 μm width and 100 nm length are prepared to form larger and smaller asymmetric SQUIDs. Subsequently, we place the gate electrodes on all the JJs to control them by the gate voltages. As depicted in Fig. 1(c), we bias an electrical current $I_1$ when measuring the larger SQUID, and measure the voltage difference $V_1$ as well. All the measurements were performed at 10 mK of the base temperature in our dilution refrigerator. In case of inducing the phase shift in JJ2 and JJL2, we added a bias current $I_2$ in the smaller SQUID.

First, we characterize the single JJ properties of JJ1 and JJ2 with the other JJs pinched off (see supplementary note 1 and figure S1). The I-V curve of JJ1 at an out-of-plane magnetic field $B$=0 mT with $V_{g1} = -1.4$ V of the gate voltage for JJ1 with the other JJs pinched off, is portrayed in Fig. 1(d), wherein the supercurrent flows and the switching current of JJ1 is 0.2 μA. The switching current of JJ1 is highly tunable as shown in Fig. 1(e). The JJ2 indicates a similar switching current and the dependence on the gate voltage $V_{g2}$ for JJ2 as shown in Fig. 1(f).

As the first step, we evaluate the CPR of the single JJ1 by measuring $V_1$ and $I_1$ of the larger asymmetric SQUID with $V_{g1} = -1.4$ V, JJ2 off, $V_{gL1} = -1.4$ V, and JJL2 off. For the evaluation, $V_1$ as a function of $I_1$ and $B$ are measured to obtain the switching current of the asymmetric SQUID. The asymmetric SQUID is utilized to evaluate the CPR by the switching current measurement because the switching current is approximately written as $I_{sc1}(\phi_{1c} + 2\pi\Phi_1(B)/\Phi_0) + I_{swL1}$ [15,25]. Here the single JJ1 CPR and the switching current of the single JJL1 are $I_{sc1}(\phi_{1c} + 2\pi\Phi_1(B)/\Phi_0)$ and $I_{swL1}$, respectively. $\phi_{1c}$ and $\Phi_0 = h/2e$ denote the constant phase difference decided by the critical current of JJL1 and the loop inductance, and flux quantum, respectively. $\Phi_1(B)$ represents the magnetic flux in the larger loop, which depends on $B$ and the loop area. Therefore, we subtract the background assigned to the Fraunhofer-type interference in JJL1 to obtain the single JJ1 CPR (see supplementary note 2 and figure S2). Consequently, the single JJ1 CPR curve, $I_{sc1}(\phi_{1c} + 2\pi\Phi_1(B)/\Phi_0)$, is obtained as shown with black circles in Fig.1 (g). We assume that the circulating supercurrent in the loop is ignored and the magnetic flux in the loops is linear to $B$ (see supplementary note 4 and figure S4). The CPR periodically oscillates and its shape is skewed from the sinusoidal function of $B$, reflecting the short ballistic nature of JJ1. These results assure that our asymmetric SQUID can be used to evaluate the JJ1 CPR.

Then the CPR of JJ1 coupled with JJ2 is studied with $(V_{g1}, V_{g2}) = (-1.4 \text{ V}, -1.45 \text{ V})$ and $(V_{gL1}, V_{gL2}) = (-1.4 \text{ V}, -1.15 \text{ V})$. These gate voltages produce similar switching currents of 0.2 μA in JJ1 and JJ2. For the sake, $V_1$ obtained as a function of $I_1$ and $B$ is measured with JJL1 and JJL2 on. The coupled JJ1 CPR curve obtained by subtracting the same background data as the single JJ1 CPR evaluation is shown as green circles in Fig.1(g). The obtained CPR curve is highly modulated from that of JJ1 with no JJ2 and is not periodic. This modulation originates from the short-range coupling between JJ1 and JJ2, which makes the CPR of JJ1 dependent on $\phi_2$. Now $B$ changes not only $\phi_1$ but also $\phi_2$ because both of the asymmetric SQUIDs are formed. Then $\phi_2$ is introduced as $2\pi\Phi_2(B)/\Phi_0$, where $\Phi_2(B)$ represents the magnetic flux in the smaller loop. As $|\Phi_2| < |\Phi_1|$ holds from the loop area relation, $\phi_1$ and $\phi_2$ evolve with $B$ at the different ratios. Therefore the $(\phi_1(B), \phi_2(B))$ trace is depicted in Fig. 1(h).

For instance, if $B$ increases from $\Phi_1 = \Phi_2 = 0$, the trace of $(\phi_1, \phi_2) = (\phi_{1c} + 2\pi\Phi_1(B)/\Phi_0, 2\pi\Phi_2(B)/\Phi_0)$ moves with $B$ as a parameter from $(\phi_{1c}, 0)$ along the solid arrows in Fig. 1(h). Reaching at $\phi_1 = \pi$, the trace is shifted to $\phi_1 = -\pi$ with the same $\phi_2$ and moves along the solid arrow. The described solid lines correspond to the trace when $\phi_1$ alters for $7 \times 2\pi$. We note that -0.9 mT < $B$ < 0.3 mT includes around 7 periods in $\phi_1$ seen in the single JJ1 CPR curve in Fig. 1(g). Therefore, the green circles in Fig. 1(g) represent $I_{sc1}(\phi_{1c} + 2\pi\Phi_1(B)/\Phi_0, 2\pi\Phi_2(B)/\Phi_0)$ and the nonperiodic dependence on $B$ indicates that the JJ1 CPR depends on $\phi_2$ via the short-range coherent coupling.

The trace lines do not fill the $(\phi_1, \phi_2)$ plane in Fig. 1(h). Therefore, only a portion of the two-dimensional CPR can be constructed from the green circles in Fig. 1(g). A way to fill the entire $(\phi_1, \phi_2)$ plane is to shift the trace lines along the vertical $\phi_2$ axis and obtain the $(\phi_{1c} + 2\pi\Phi_1(B)/\Phi_0, 2\pi\Phi_2(B)/\Phi_0 + \Delta\phi_2)$ traces. Here $\Delta\phi_2$ represents the shift of $\phi_2$. For the sake, we add the bias current $I_2$ in the smaller SQUID because the finite supercurrent in the asymmetric SQUID shifts the phase differences of JJs following their CPRs and the SC loop inductance. To evaluate $\Delta\phi_2$ induced by $I_2$, we measure the JJ1 with $(V_{g1}, V_{g2}) = (-1.4 \text{ V}, -1.45 \text{ V})$, JJL1 off, and $V_{gL2} = -1.15$ V. Under this condition, the switching current of JJ1 depends on $\phi_2$ due to the nonlocal Josephson effect derived from the coherent coupling [28,31].

$I_{sw1}$ is displayed in Fig. 2(a) as a function of $B$ for several $I_2$ values between -600 and 600 nA. $I_{sw1}$ oscillates with $B$, originating from the coherent coupling between JJ1 and JJ2. The oscillation pattern gradually shifts along the $B$ axis as $I_2$ varied. This explicitly implies that $\Delta\phi_2$ is introduced by $I_2$ and the switching current is described as $I_{sw1}(2\pi\Phi_2(B)/\Phi_0 + \Delta\phi_2(I_2))$. As the single oscillation period corresponds to $2\pi$, $\Delta\phi_2(I_2)/2\pi$ is estimated from the $I_{sw1}(2\pi\Phi_2(B)/\Phi_0 + \Delta\phi_2(I_2))$ curve as a ratio of the shift along the $B$ axis from the curve of $I_2 = 0$ nA to the single oscillation period in $B$. The estimated $\Delta\phi_2$ versus $I_2$ is portrayed in Fig. 2(b). In the range of -600 nA $\leq I_2 \leq$ 600 nA, $\Delta\phi_2$ is tuned by $\sim 0.4\pi$.

Subsequently, we measure the CPR of JJ1 with JJ2 on by varying $I_2$. The CPR curves are presented as a function of $B$ in Fig. 2(c). The result for $I_2 = 0$ nA corresponds to the green data in Fig. 1(g). The curve shape gradually varies with $I_2$, reflecting $I_{sc1}(\phi_{1c} + 2\pi\Phi_1(B)/\Phi_0, 2\pi\Phi_2(B)/\Phi_0 + \Delta\phi_2)$ (see supplementary note 3 and figure S3). Thus, we obtain the necessary CPR dataset to construct $I_{sc1}(\phi_1, \phi_2)$ in $-\pi \leq \phi_1 \leq \pi$ and $-\pi \leq \phi_2 \leq \pi$ (see supplementary note 3 and figure S3). For clarity, we represent $I_{sc1}(\phi_1, \phi_2)$ in $-2\pi \leq \phi_1 \leq 2\pi$ and $-2\pi \leq \phi_2 \leq 2\pi$ in Fig. 3(a) by pasting several $I_{sc1}(\phi_1, \phi_2)$ data with $\pm 2\pi$ shift along the $\phi_1$ or $\phi_2$ axes. The constructed CPR evidently depends on both $\phi_1$ and $\phi_2$ and looks point-symmetric to $\phi_1 = \phi_2 = 0$ as expected from the time-reversal relation of $I_{sc1}(\phi_1, \phi_2) = -I_{sc1}(-\phi_1, -\phi_2)$ for the coupled JJs.

We numerically calculate the CPR of two coupled planar JJs. We use the tight-binding model to evaluate the energies of the Andreev bound states formed in the coherently coupled JJs and calculate the supercurrent in JJ1 by the differential of the total energies by $\phi_1$. The obtained numerical result is presented in Fig. 3(b) (see supplementary note 5 and figure S5), which gives an excellent agreement with the experimental results. This agreement supports that the evaluated data exhibit the desired two-dimensional CPR of $I_{sc1}(\phi_1, \phi_2)$.

To discuss the feature of the φ junction, we focus on $\phi_1$ yielding $I_{sc1}(\phi_1, \phi_2) = 0$ nA because no supercurrent flows at the ground state of the JJ. We plot $\phi_1$ with $I_{sc1}(\phi_1, \phi_2) = 0$ nA as a function of $\phi_2$ highlighted by a purple line in Fig. 3 (a)). Here we only indicate $\phi_1$ continuously connecting to $\phi_1 = \phi_2 = 0$ and ignore the other $I_{sc1}(\phi_1, \phi_2) = 0$ nA around $\phi_1 = \pi$ because the system is time-reversal and spatial-inversion invariant at $\phi_2 = 0$, resulting in $\phi_1 = 0$ as the ground state of JJ1. The ground state gradually progresses along the purple line away from $\phi_1 = 0$ as $\phi_2$ alters from $\phi_2 = 0$, indicating the φ junction of JJ1. The purple line oscillates with $\phi_2$ period of $2\pi$, which clearly proves that the φ junction originating from the coherent coupling can be tuned by a nonlocal phase difference.

To find the spontaneous supercurrent, i.e. the supercurrent with no local phase difference of $\phi_1=0$, we plot the line profiles of Fig. 3(a) at $\phi_2=0$, $\pm\pi/2$, and $\pi$ in Fig. 3(c). It is clear to see the curves at $\phi_2=\pm\pi/2$ holding a finite supercurrent at $\phi_1=0$. Then, for confirming the gate tunability of the spontaneous supercurrent, we plot the line profile of Fig. 3(a) at $\phi_1=0$ in Fig. 3(d), which represent $I_{sc1}(0, \phi_2)$. $I_{sc1}(0, \phi_2)$ oscillates as a function of $\phi_2$. This means the finite spontaneous supercurrent in the coupled JJs controlled by the nonlocal phase difference. The spontaneous supercurrent as a function of $\phi_2$ is approximately an odd function as expected from the time-reversal relation of $I_{sc1}(0, \phi_2) = -I_{sc1}(0, -\phi_2)$.

Finally, we explore $V_{g1}$ and $V_{g2}$ dependence of the two-dimensional CPR of JJ1. The evaluated JJ1 CPR with several $(V_{g1}, V_{g2})$ are summarized in Fig. 4 (see supplementary note 4 and Figure S4). For example, the panel at the upper-left corner shows $I_{sc1}(\phi_1, \phi_2)$ at $(V_{g1}, V_{g2}) = (-1.4 \text{ V}, -1.95 \text{ V})$. The $I_{sc1}$ results measured for the same $V_{g1}$ but different $V_{g2}$ values are shown on the same color scales to the right. Figure 4 exhibits that the AJE behavior (namely $\phi_2$ dependence of $I_{sc1}$) decreases as $V_{g2}$ is made more negative while the φ junction behavior is less modulated by $V_{g1}$. This gate dependence can be assigned to the relation of the numbers of ABSs in JJ1 and JJ2. In the case that the number of ABSs in JJ1 is smaller than that in JJ2, all the ABSs in JJ1 are coupled with the ABSs in JJ2. Therefore, the AJE behavior can remain evident. On the other hand, in the opposite case, some ABSs in JJ1 are not coupled with the ABSs in JJ2, resulting in a less clear AJE behavior. This gate voltage dependence can be reproduced in our numerical calculation (see supplementary note 6 and figure S6). Then the gate dependence supports that the AJE in JJ1 emerges from the coherent coupling with JJ2. We note that the AJE appears even in $V_{g1} = -1.7$ V where $I_{sc1}$ is small or in a 10 nA order. This means that the AJE behavior in JJ1 is not an artifact from our evaluation method in which we assume $\Phi_1$ and $\Phi_2$ linear to $B$ and ignore the inductance term related to the circulating supercurrent in the SC loop because the small supercurrent makes the inductance effect smaller (see supplementary note 4 and figure S4).

In our consideration, we do not include the spin-orbit interactions which the InAs quantum well holds. At least, the obtained AJE can be explained only by the coherent coupling and it is difficult to discuss the roles of the spin-orbit interactions in these results. To understand such physics, it may be necessary to study the SC transport in the coupled JJs with the in-plane magnetic fields because the in-plane magnetic fields produce the various exotic SC phenomena related to the spin-orbit interactions in the single JJs such as the SC phase batteries, SC diodes, and Majorana zero modes.

In conclusion, we construct a two-dimensional CPR of a JJ coherently coupled to another JJ using asymmetric SQUIDs. From the CPR, we demonstrate the spontaneous supercurrent and φ junction controlled by the nonlocal phase difference, indicating the phase-tunable AJE. The obtained AJE contributes to the development of novel SC devices such as SC phase batteries and SC diodes. We note our method for the two-dimensional CPR evaluation will be applicable to multiterminal JJs as well, where a single normal metal is coupled to several SC electrodes [42–48].

## Figures

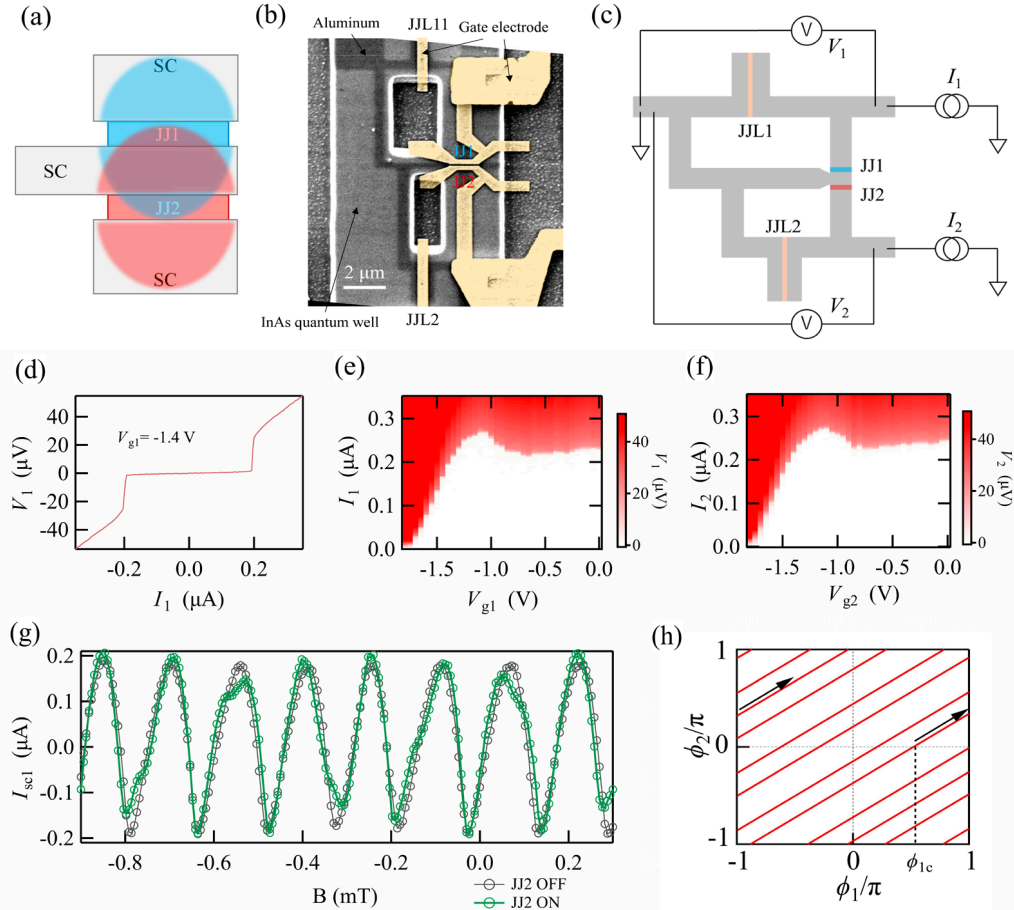

**Fig. 1: Concept and device**

(a) Conceptual image of coupling between two JJs. Overlapping of wavefunctions of Andreev bound states in the shared SC electrode connects the two JJs to form Andreev molecule states.

(b) An SEM image of our device. The black region represents the aluminum electrodes. The respective JJs are covered by the gate electrodes highlighted with yellow.

(c) A schematic of our device and measurement circuit.

(d) A typical I-V curve of the single JJ1 at $V_{g1} = -1.4$ V with the other JJs off at $B = 0$ mT.

(e) $V_1$ as a function of $I_1$ and $V_{g1}$ at $B = 0$ mT to characterize the gate dependence of the single JJ1 with the other JJs off. The supercurrent of JJ1 is highly tunable in the $V_{g1}$ range of $-1.9\text{ V} < V_{g1} < -1.3\text{ V}$. Then the CPR of JJ1 is studied in this range.

(f) $V_2$ as a function of $I_2$ and $V_{g2}$ at $B = 0$ mT to characterize the gate dependence of the single JJ2. The supercurrent of JJ2 is highly tunable in $-2.0\text{ V} < V_{g1} < -1.4\text{ V}$.

(g) The supercurrent in the single JJ1 (the black circles) and the JJ1 coupled to JJ2 (the green circles) as a function of $B$ evaluated from the switching current measurement of the larger asymmetric SQUID. The single JJ1 results periodically oscillate to $B$ while the coupled JJ1 results are not periodic due to modulation from the coherent coupling.

(h) The $(\phi_1, \phi_2)$ trace as $B$ is swept in the case that all the JJs are on. From the measurement of the larger SQUID with $I_2 = 0$ nA, the JJ1 CPR on the trace lines can be obtained.

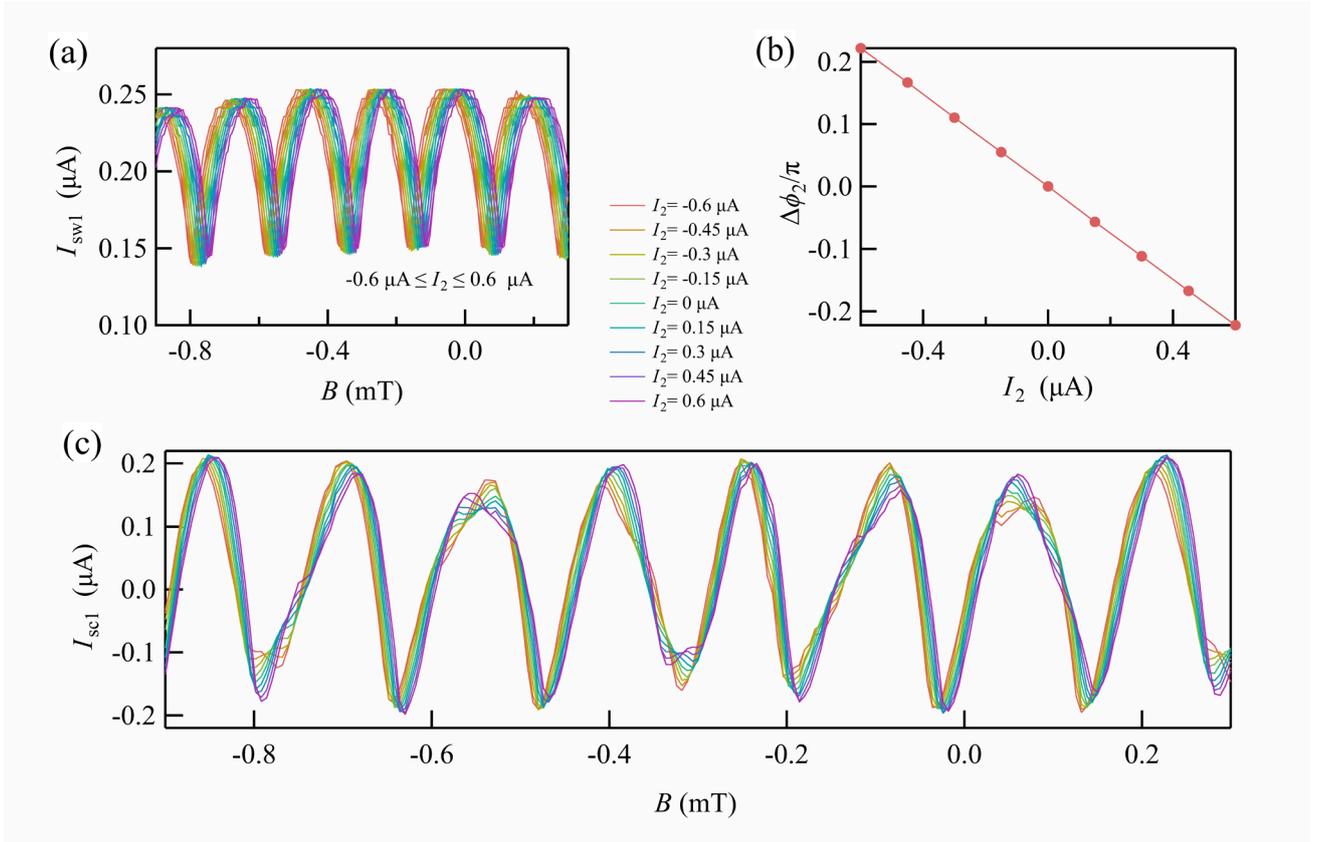

**Fig. 2: Necessary dataset to obtain the two-dimensional CPR of JJ1 coupled to JJ2**

(a) Switching current of JJ1 ($I_{sw1}$) as a function of $B$ at $-0.6\ \mu A \leq I_2 \leq 0.6\ \mu A$ when JJ2 and JJL2 are on. The oscillation of switching current originates from the nonlocal Josephson effect derived from the coupling of JJ1 and JJ2. The curves shift along the $B$ axis as $I_2$ varies, implying that the phase shift of JJ2 ($\Delta\phi_2$) is induced.

(b) Evaluated $\Delta\phi_2$ as a function of $I_2$ from (a). $\Delta\phi_2$ is tunable with $I_2$ in the ramge of $-0.2\pi < \Delta\phi_2 < 0.2\pi$.

(c) The obtained JJ1 supercurrent as a function of $B$ at $-0.6\ \mu A \leq I_2 \leq 0.6\ \mu A$. Line colors specifying $I_2$ are consistent with those in (a). The curve shapes are modulated by changing $I_2$.

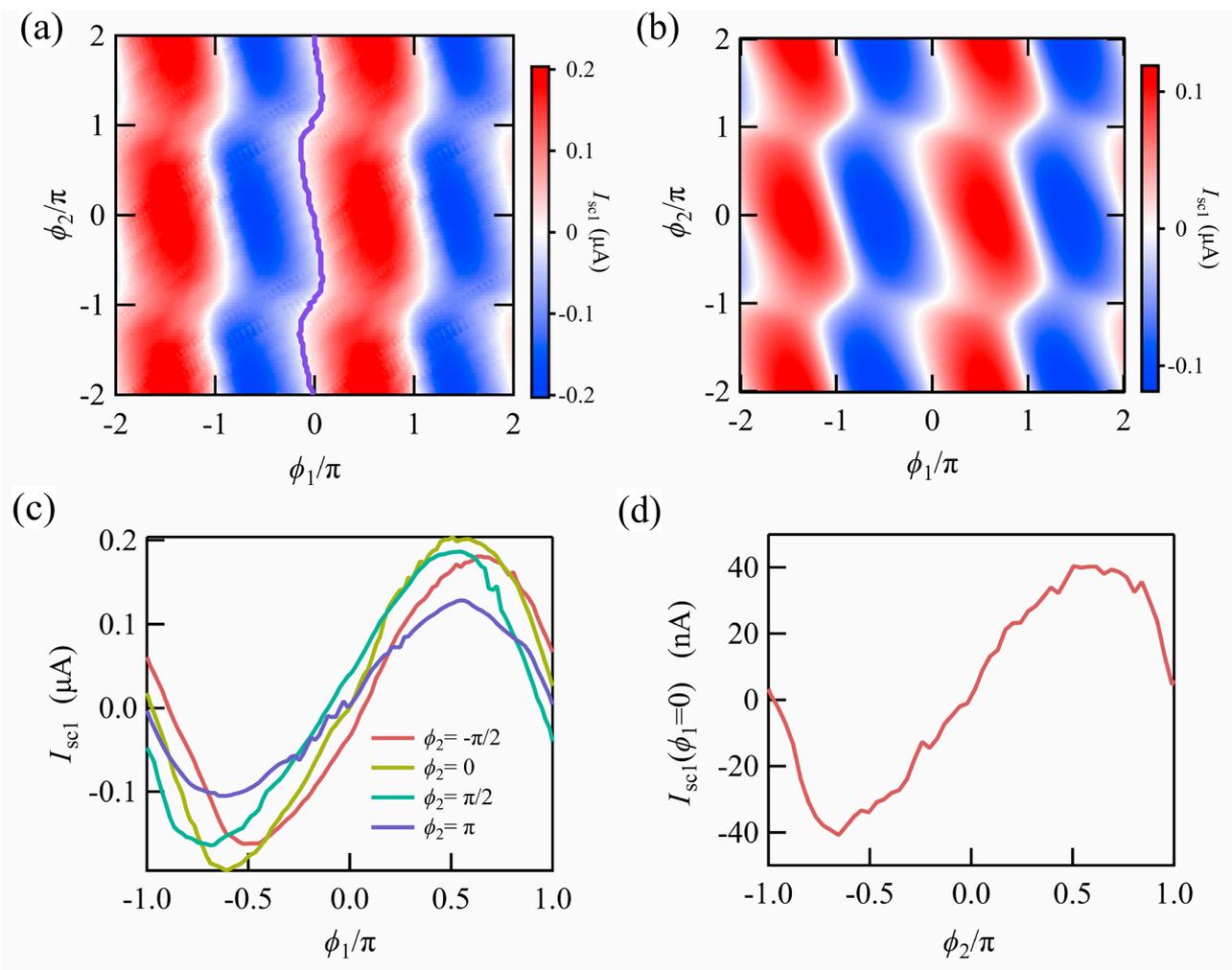

**Fig. 3: The CPR of JJ1 coupled to JJ2 and the phase-tunable anomalous Josephson effect**
(a) The obtained two-dimensional CPR of JJ1 coupled to JJ2 is shown. The JJ1 supercurrent depends not only on $\phi_1$ but also on $\phi_2$, which means that the coherent coupling of JJ1 and JJ2 produces the nonlocal Josephson effect. A purple line on $I_{sc1}(\phi_1, \phi_2) = 0$ nA is extended from $(\phi_1, \phi_2) = (0,0)$, exhibiting the evolution of $\phi_1$ for the ground state of JJ1 with $\phi_2$. Therefore, $\phi_1 \neq 0$ on the purple line indicates that the φ junction is formed and tunable by $\phi_2$.
(b) Numerically calculated CPR of JJ1 coupled to JJ2 using the tight-binding model. This gives a good agreement with the experimental result in (a).
(c) Line profiles at $\phi_1 = 0, \pm\pi/2, \pi$ in (a) are shown.
(d) A line profile at $\phi_1 = 0$ in (a) is shown. The finite supercurrent flowing in JJ1 with $\phi_1 = 0$ and the spontaneous supercurrent is tunable with $\phi_2$.

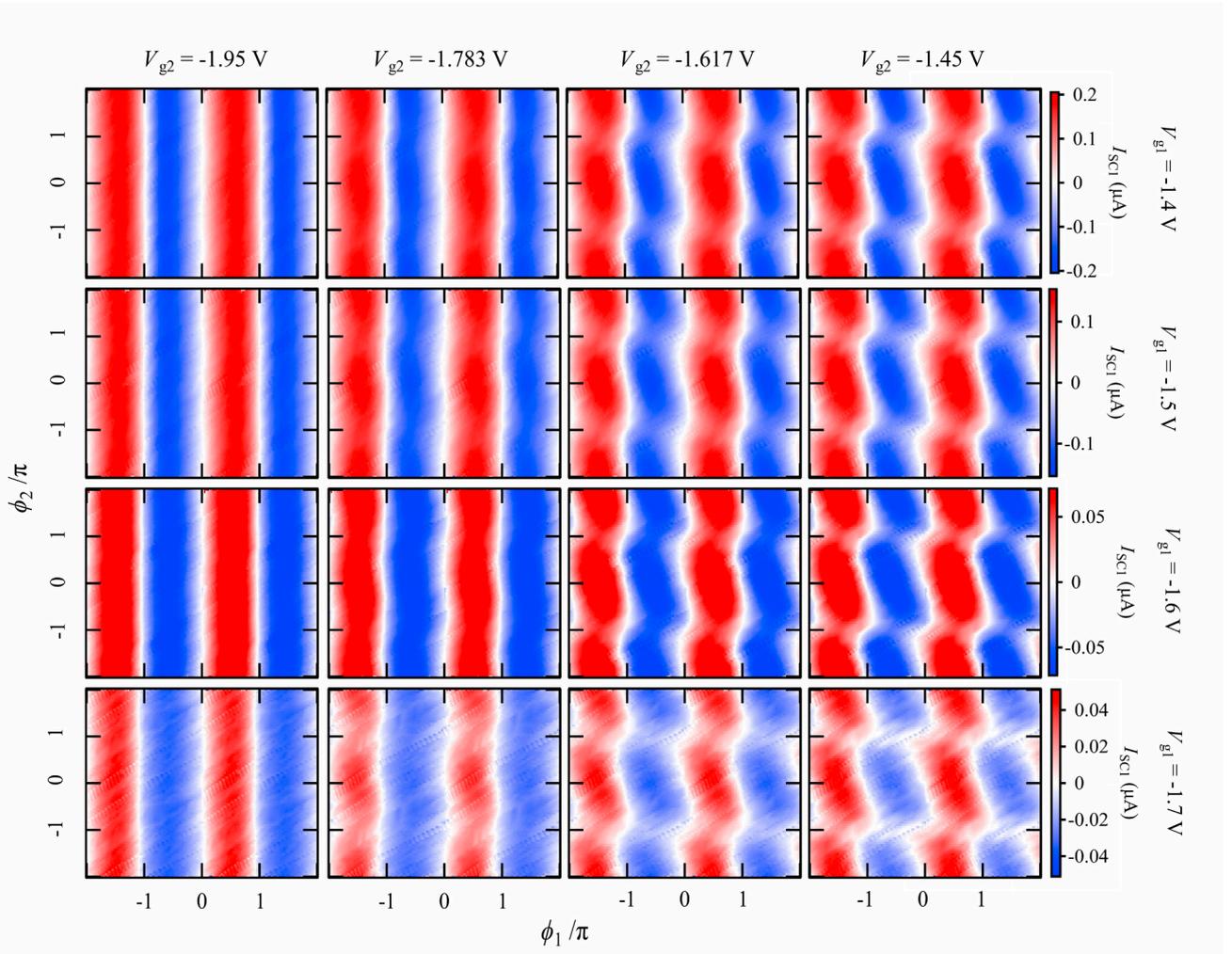

**Fig. 4: Gate voltage dependence of the two-dimensional CPR of JJ1**

The JJ1 CPR $I_{sc1}(\phi_1, \phi_2)$ results obtained at several sets of $(V_{g1}, V_{g2})$ are shown. The panels in the same row (column) are obtained at the same $V_{g1}$ ($V_{g2}$), labeled on the right side (upside). In addition, the same row images are depicted with the same color scales placed on the right side. These results indicate that the dependence on $\phi_2$ becomes weaker as $V_{g2}$ becomes more negative while it is almost unchanged though $V_{g1}$ is varied.

## Methods

*Sample growth*

The wafer structure has been grown via molecular beam epitaxy on a semi-insulating InP substrate. The stack materials from bottom to top are a 100 nm $In_{0.52}Al_{0.48}As$ buffer, a 5 period 2.5 nm $In_{0.53}Ga_{0.47}As$/2.5 nm $In_{0.52}Al_{0.48}As$ superlattice, a 1 μm thick metamorphic graded buffer stepped from $In_{0.52}Al_{0.48}As$ to $In_{0.84}Al_{0.16}As$, a 33 nm graded $In_{0.84}Al_{0.16}As$ to $In_{0.81}Al_{0.19}As$ layer, a 25 nm $In_{0.81}Al_{0.19}As$ layer, a 4 nm $In_{0.81}Ga_{0.19}As$ lower barrier, a 5 nm InAs quantum well, a 10 nm $In_{0.81}Ga_{0.19}As$ top barrier, two monolayers of GaAs, and finally an 8.7 nm layer of epitaxial Al. The top Al layer has been grown in the same chamber without breaking the vacuum. The two-dimensional electron gas (2DEG) accumulates in the InAs quantum well.

*Device Fabrication*

In this study, conventional electron beam lithography was used to fabricate JJs. We etched out the aluminum film using the type-D etchant after we defined the mesa of the InAs quantum well with 1:1:8 of $H_3PO_4:H_2O_2:H_2O$ etchant. Subsequently, we grew a 30-nm-thick $Al_2O_3$ film by atomic layer deposition and deposited Ti and Au for the gate electrodes.

*Measurement*

For the measurement of the switching current, we measured the I-V curves of JJ1 for various conditions. When switching JJL1 or JJL2 off to measure the single or coupled JJ1, we set $V_{gL1} \leq -1.9$ V or $V_{gL2} \leq -2.0$ V. When switching JJL1 or JJL2 on to form the asymmetric SQUIDs, we set $V_{gL1} = -1.4$ V or $V_{gL2} = -1.15$ V. The switching currents of JJL1 and JJL2 at $V_{gL1} = -1.4$ V and $V_{gL2} = -1.15$ V are 0.8 μA and 1.0 μA, respectively. When pinching off JJL1 and JJL2, we set $V_{gL1}\left(V_{gL2}\right) = -4$ V.

## Acknowledgments

This work was partially supported by a JSPS Grant-in-Aid for Scientific Research (S) (Grant No. JP19H05610), JST PRESTO (grant no. JPMJPR18L8), JSPS Grant-in-Aid for Early-Career Scientists (Grant No. 18K13484), Advanced Technology Institute Research Grants, and the Ozawa-Yoshikawa Memorial Electronics Research Foundation.

**Author contributions**

S.M. designed the experiments. T.L., S.G., G.C.G, and M.J.M grew wafers to form InAs 2DEG quantum wells covered with epitaxial aluminum. S.M. fabricated the devices. S.M. and T.I. performed measurements. S.M., T.I., Y.S., and S.T. analyzed the data. T.Y. performed numerical calculations. S.T. supervised the study.

**Competing interest statement**

The authors declare no competing interests.

**Supplementary information**

Supplementary Notes 1–5 and Supplementary Figures S1–5.

**Corresponding authors**

Correspondence and requests should be addressed to Sadashige Matsuo (sadashige.matsuo@riken.jp) and Seigo Tarucha (tarucha@riken.jp).

# Supplementary Information for

# Engineering of anomalous Josephson effect in coherently coupled Josephson junctions

**Supplementary Note 1: Gate control of independent JJs**

We independently control the JJs named JJ1, JJ2, JJL1, and JJL2. The gate voltages on the respective JJs are denoted as $V_{g1}$, $V_{g2}$, $V_{gL1}$, and $V_{gL2}$. The gate control of the I − V curve in the single JJL1 and JJL2 with the other JJs pinched off is portrayed in Fig. S1 for JJL1 in (a) and JJL2 in (b), respectively. From the results, when we switch off the JJs, we applied the gate voltages of -4 V.

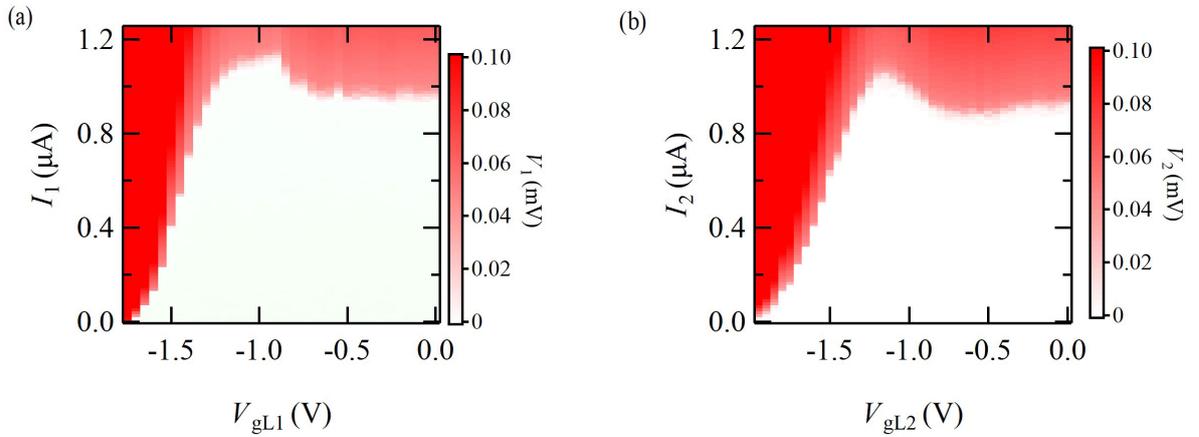

Fig. S1 The voltage differences on the single JJL1, and JJL2 with the other JJs pinched off as a function of the bias currents and the gate voltages in (a), and (b), respectively.

**Supplementary Note 2: Subtraction of the background to obtain the CPR**

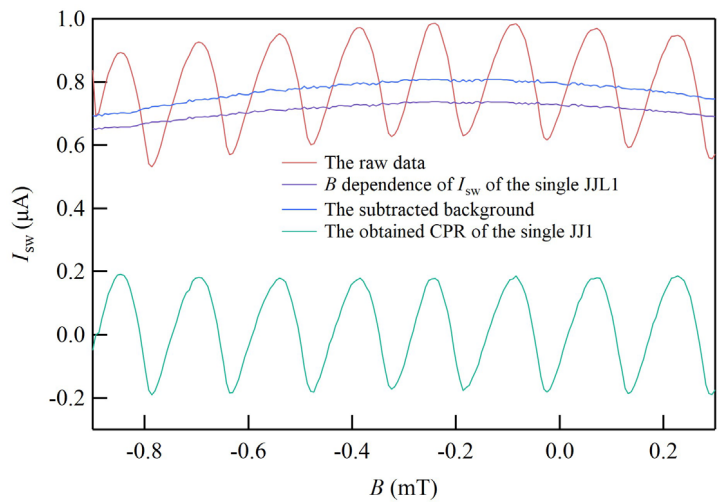

Fig. S2 The raw data obtained by the measurement of the SQUID of JJ1 and JJL1, the $B$ dependence of $I_{sw}$ of the single JJL1 with the others pinched off, the background data, and the obtained CPR data of the single JJ1 as a function of $B$ in red, purple, blue, and green, respectively.

We measure the asymmetric SQUID to obtain the CPR. To obtain the CPR of JJ1 with JJ2 on, we measure the $B$ dependence of supercurrent in JJL1 and subtract it from the raw data displayed with red in Fig. S2. Instead of using the digital low-path filter, we take a strategy to use the single JJL1 data as a background. Thus, we derive the $B$ dependence of the switching current in the single JJL1 (with the other JJs pinched off) as exhibited with the purple in Fig. S2. This dependence originates from the Fraunhofer-type interference of the single JJL1. We use the constant multiple (1.34) of the purple curve to remove the curved component of the background. After that, we remove the constant from the data to make the maximum $I_{sw}$ equal to the absolute value of the minimum $I_{sw}$. As a result, the subtracted background and the obtained CPR of the single JJ1 are depicted as blue and green in Fig. S2, respectively.

When subtracting the background from the raw data with JJ2 on, we use the same background in Fig. S2.

**Supplementary Note 3: Evaluation of $\phi_1$ and $\phi_2$**

The two-dimensional CPR can be obtained by converting the data as a function of $B$ to $\phi_1$ and $\phi_2$, which indicates the evaluation of $\phi_1(B)$ and $\phi_2(B)$. We need such a coversion procedure in the results of Fig. 1(g) or 2(c). In this procedure, we ignore the effect of the inductance of the superconducting arms (see Supplementary Note 4). First, we show the evaluation of $\phi_1(B)$ from the switching current data of JJ1 and JJL1 asymmetric SQUID with JJ2 pinched off. In this case, the grey curve in Fig. 1(g) reflects the CPR of the short ballistic Josephson junctions represented as $I_{sc1}(\phi_1) = \frac{A\tau \sin\phi_1}{\sqrt{1-\tau(\sin\frac{\phi_1}{2})^2}}$, where $A$ and $\tau$ are the coefficient and transmission of the junction, respectively. Therefore, we can fix the $B$ points providing $\phi_1 = 0$ (mod $2\pi$) in the JJ1 CPR data as a function of $B$ as the points at which the CPR becomes 0 and the slope of the tangent line is positive. The $B$ points giving $\phi_1 = 0, \pi$ (mod $2\pi$) are shown in Fig. S3(a). For the conversion from $B$ to $\phi_1$, we linearly connect the points next to each other and use the curve. We use the same curve for the $B$ to $\phi_1$ conversion in the case of JJ2 pinched on.

Then we move to conversion from $B$ to $\phi_2$. For this sake, we measure the asymmetric SQUID of JJ2 and JJL2 with JJ1 and JJL1 pinched off. The JJ2 CPR data from the positive switching current of the SQUID is represented as $I_{sc2+} = I_{sc2}(\phi_{2c} + 2\pi\Phi_2(B)/\Phi_0)$, where $\phi_{2c}$ and $\Phi_0 = h/2e$ denote the constant phase difference and flux quantum, respectively. As well as the positive case, the JJ2 CPR data from the negative switching current can be obtained as $I_{sc2-} = I_{sc2}(-\phi_{2c} + 2\pi\Phi_2(B)/\Phi_0)$. Therefore, we can decide the $B$ points giving $2\pi\Phi_2(B)/\Phi_0 = 0, \pi$ (mod $2\pi$) as points at which $I_{sc2+} + I_{sc2-}$ shown in Fig. S3(b) becomes $I_{sc2}(\phi_{2c}) + I_{sc2}(-\phi_{2c}) = 0$ because of $I_{sc2}(-\phi_{2c}) = -I_{sc2}(\phi_{2c})$ forced from the time-reversal symmetry. Then we decide $2\pi\Phi_2(B)/\Phi_0 = 0$ (mod $2\pi$) from the measurement of the JJ1 switching current with JJL1 pinched off, JJ2 on and JJL2 on shown in Fig. 2(a) with $I_2 = 0$ A. From the previous reports, the JJ1 switching current oscillation becomes maximum and minimum around $2\pi\Phi_2(B)/\Phi_0 = 0$ and $\pi$ (mod $2\pi$), respectively. Therefore, we fix the $B$ points giving $2\pi\Phi_2(B)/\Phi_0 = 0$ (mod $2\pi$). The evaluated points are summarized in Fig. S3(c) and we make the $B$ to $\phi_2$ conversion curve by connecting the points next to each other linearly.

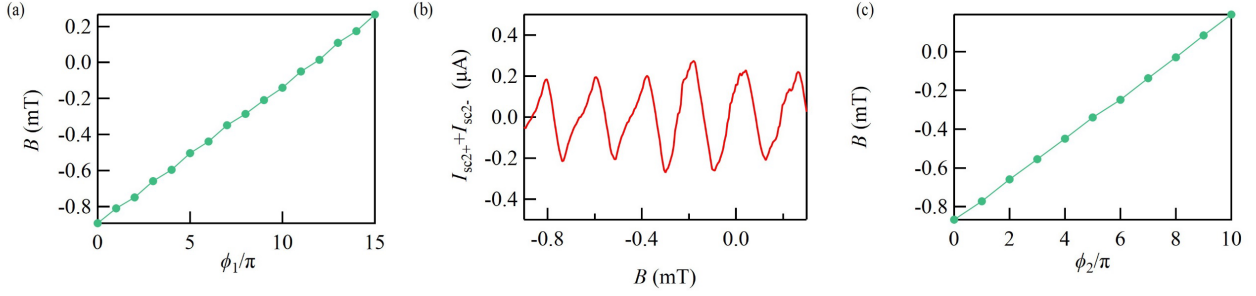

Fig. S3 (a)The conversion curve for $\phi_1$ from $B$ derived from the evaluated CPR of the single JJ1. (b)The sum of the positive and negative switching currents obtained by the measurement of the SQUID of JJ2 and JJL2 with JJ1 and JJL1 pinched off. From this curve, we can decide the $B$ points corresponding to $\phi_2 = 0$ (mod $\pi$) as shown in (c).

**Supplementary Note 4: Inductance effect**

There are the effects of the finite inductance of the SQUID superconducting arms. In our analysis, we ignore the inductance effects invoked by the supercurrent in the JJ1 and JJ2 but consider the effects by the supercurrent in the JJL1 and JJL2 as written below. The inductance measurement appears through $\phi_1 - \phi_{L1} = \{2\pi\Phi_1(B) - ((L_1 + L_{L1})I_{cw1} - (L_1 - L_{L1})I_1)\}/\Phi_0$ and $\phi_2 - \phi_{L2} = \{2\pi\Phi_2(B) - (L_2I_2 - L_{L2}I_{L2})\}/\Phi_0$. These equations lead to two effects:

1. The skewed conversion curve for $\phi_1$ from $B$
2. The phase shift due to the bias current in the SQUID.

The effect 1 originates from the term including $I_{cw1}$. We ignore this as found in supplementary note 3 to obtain $\phi_1$ and $\phi_2$ from the conversion. In the case that the single JJ1 CPR evaluation with JJ2 off, this can be included by constructing the full model using the analytical form of the CPR in the short-ballistic JJ. On the other hand, this cannot work in our coupled JJ case because no analytical solution of the two-dimensional CPR is known at present and then it is impossible to assume the circulating supercurrent in the loop. We note that it is possible to conclude that this approximation (ignoring the inductance effect 1) does not produce the observed AJE and φ junction. The smaller $I_{cw1}$ produces the smaller effect from the term $(L_1 + L_{L1})I_{cw1}$ and therefore, if the AJE originates from this effect, the smaller $I_{cw1}$ decreases or vanishes the AJE. However, the AJE and φ junction can explicitly be found even with the smaller switching current of JJ1 (namely smaller $I_{cw1}$) as shown in the panels of $(V_{g1}, V_{g2}) = (-1.7\text{ V}, -1.45\text{ V})$ and $(-1.7\text{ V}, -1.617\text{ V})$ in Fig. 4.

The effect 2 is due to the inductance, which appears when making $\phi_2$ shift by $I_2$ in Fig. 2(a). The obtained $\Delta\phi_2$ includes the shift invoked by the $(L_2I_2 - L_{L2}I_{L2})$ term and also the change of $\phi_{L2}$ by the bias current $I_2$. This effect also appears when we measure the switching current of SQUID. When we decide the conversion curve for $\phi_1$ from the measurement of SQUID of JJ1 and JJL1 with JJ2 and JJL2 off, the inductance effect 2 is included.

On the other hand, the situation for the results in Fig. 2(c) with the bias current $I_2$, the effect 2 becomes complicated because not only the bias currents of $I_1$ but also $I_2$ can shift the phase differences through the inductance effect. This inductance effect invoked by $I_2$ makes a change in the $B$ to $\phi_1$ conversion curve which is obtained from the measurement of SQUID with $I_2 = 0$ nA. Therefore, we try to evaluate this effect approximately from the additional experiments as below.

We assume that almost all the bias current $I_2$ flows through JJL2 and into the GND because the JJ2 and JJ1 are sufficiently smaller than JJL2. With this assumption, we can evaluate the effect of $I_2$ on the conversion curve for $\phi_1$ by measuring the SQUID of JJ1 and JJL1 with JJ2 off and JJL2 on, flowing $I_2$. Figure S4(a) indicates the JJ1 CPR evaluated from the SQUID switching current with several $I_2$. As expected, $I_2$ shifts the oscillation through the inductance effect 2 ($I_2$ in the inductance of the SC arm 1 makes the phase shift of the JJ1). From these results, we can obtain the conversion curves for $\phi_1$ at several $I_2$ in Fig. S4(b) which includes the simplified inductance effect invoked by $I_2$. To obtain the two-dimensional CPR results from Fig. 2(c), we use these conversion curves.

To obtain the results in Fig. 4, we use the same conversion curves for $\phi_1$ and $\phi_2$ at several $I_2$ in Fig. S4(b) and (c) as those used to obtain Fig. 3(a). This is based on an assumption that almost all the bias current flows in the larger JJs (JJL1 and JJL2). When $V_{g2} = -1.95$ V, JJ2 is almost pinched off. Therefore, in this situation, $\phi_2$ is not defined. However, we indicate the results in the ($\phi_1$,$\phi_2$) plane in Fig. 4 obtained with the conversion curves for $\phi_2$ in Fig. S4(c) to confirm that our analysis scheme produces the expected results (no $\phi_2$ dependence) even with JJ2 pinched off.

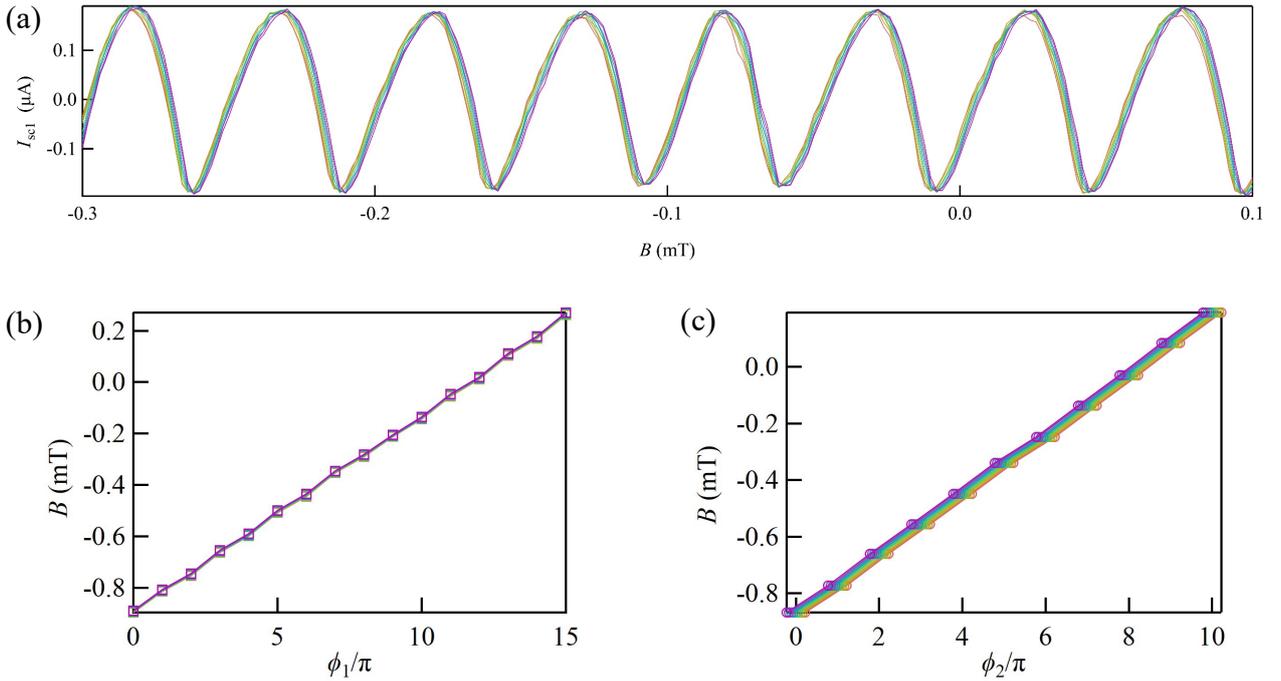

Fig. S4 (a) The obtained CPR for the single JJ1 with the several $I_2$ by the measurement of the SQUID of JJ1 and JJL1 with JJ2 off and JJL2 on. The correspondence of the curve colors and the $I_2$ values is the same as that indicated in Fig. 2. (b) The conversion curves for $\phi_1$ at the several $I_2$. (c) The conversion curves for $\phi_1$ at the several $I_2$. Horizontal shift of these curves corresponds to $\Delta\phi_2$ induced by $I_2$.

There are several small jumps of the data in the evaluated JJ1 two-dimensional CPR in Fig. 3(a) even with the corrections for the inductance effect. This may be caused by the simplification of our evaluation of these inductance effects. In reality, the conversion curves for $\phi_1$ and $\phi_2$ are nonlinear due to the inductance effect 1 and some of $I_2$ flow in JJ2 and JJ1 in the measurement of the SQUID to obtain Fig. 2(c). For a more accurate evaluation of the two-dimensional JJ1 CPR, it is demanded to treat these inductance effects. For example, our CPR results do not indicate the Josephson diode effect which has already been observed and numerically reproduced. The Josephson diode effect should appear as a difference between the absolute values of the maximum and minimum supercurrent in the JJ1 CPR as a function of $\phi_1$ with the fixed $\phi_2$. The experimentally observed Josephson diode effect in similar devices exhibits around 10 nA difference in the absolute values of the maximum and minimum switching currents. Our evaluated CPR includes the data jumps of ~10 nA. Therefore, it is impossible to discuss the Josephson diode effect in the JJ1 CPR. Further experimental studies with the optimized device structure and theoretical efforts are demanded to establish the physics of the coupled JJ CPR.

**Supplementary Note 5: Numerical calculation model**

A numerical calculation conducted in this study is explained herein. The numerical calculation is based on a tight-binding model that discretizes the real space on a square lattice. The free energy and CPR of the coupled JJs were calculated from the energy spectra of the Andreev bound states in the model. We considered the device structure presented in Fig. 1 in the main text as two coupled JJs. In addition to the normal regions governing the properties of Andreev bound states—the Josephson current, the spatial structure of the superconducting regions induced by the proximity effect of the superconducting metals is intrinsically considered in the proposed model. Moreover, the wavefunctions of the Andreev bound states penetrate the superconducting regions with exponential decay. Therefore, under a finite length of the central superconducting region, the coherent coupling between the two JJs is intrinsically introduced in the current calculation, and the CPR is modulated by the coherent coupling. However, this coupling is weakened by impurities, a large distance between the JJs, and electronic mismatches at the superconductor/normal metal interfaces. Therefore, the present experimental scheme is favorable for the coherent coupling of JJs.

The structure of the two coupled JJs displayed in Fig. 1(a) in the main text is described in the two-dimensional tight-binding model in Fig. S5. Furthermore, several junction lengths are characterized as $L_W, L_C, L_N$, and $L_S$ that denote the widths of JJ1 and JJ2, the width of the central superconductor, lengths of JJ1 and JJ2, and lengths of upper and lower superconductors, respectively. In the tight-binding model, $L_W = (N_W + 1)a$, $L_C = (N_C + 1)a$, $L_N = N_N a$, and $L_S = (N_S + 1)a$ with the lattice constant a. For the three superconducting regions (yellow shadowed regions), the strengths of the pair potentials, $\Delta_0$, were constant and equivalent to each other. For the two normal regions (white region), the Fermi energy $E_F$ provides the number of conduction channels, and consequently, the number of Andreev levels. In the model, we introduce random on-site potentials $V_{\text{imp}}(i,j)$ with a uniform distribution ranging from $\frac{W_0}{2} \geq V_{\text{imp}}(i,j) \geq -\frac{W_0}{2}$ to describe the influence of impurities. More specifically, the mean free path is estimated as $l_{\text{mfp}} = \frac{6\lambda_F^3}{\pi^3 a^2}\left(\frac{E_F'}{W_0}\right)^2$ with Fermi wavelength $\lambda_F$ and $E_F' = E_F - E_1$. $E_1$ denotes the band-edge energy of the lowest subband. In addition, we consider potential barriers in the normal regions of JJ1 and JJ2, which are tuned by gate voltages. We assume $V_\alpha(i,j) = \frac{V_{g\alpha}}{2}\left\{1-\cos\left(\frac{2\pi(j-j_\alpha)}{N_N}\right)\right\}$ for $\alpha=1,2$ only in the normal regions. Here, $V_{g1}$ and $V_{g2}$ denote the potential heights for JJ1 and JJ2, respectively. $j_1$ and $j_2$ are boundary positions of the central superconducting region with JJ1 and JJ2, respectively.

The parameters of the tight-binding model can be stated as follows: $N_W = 20, N_C = 12, N_N = 6$, and $N_S = 40$ with $a = 20$ nm for the length scale, $\frac{\Delta_0}{t} = 0.03, \frac{E_F}{t} = 0.9$, and $\frac{W_0}{t} = 2$ with $t = \frac{\hbar^2}{2m^*a^2} \simeq 4.141$ meV for the energy scale. $m^*=0.023m_e$ denotes the effective mass of the conduction band in InAs. Thereafter, the number of conduction channels by width $L_W$ is $N_{ch} = 6$, and the mean free path is $L_{mfp} \simeq 217$ nm.

The energy levels of the Andreev bound states, $E_n(\phi_1, \phi_2)$, are calculated using a numerical diagonalization of the Bogoliubov de-Gennes Hamiltonian on the tight-binding model. Moreover, the free energy at zero temperature is $E_J(\phi_1, \phi_2) = \Sigma_n E_n(\phi_1, \phi_2)$. The summation is conducted only for the positive levels, and the CPR is evaluated as

$$I_\alpha(\phi_1, \phi_2) = -\frac{e}{\hbar}\frac{\partial E_J(\phi_1, \phi_2)}{\partial \phi_\alpha}$$

with $\alpha = 1,2$. Therefore, the typical scale of the Josephson current is $\frac{e\Delta_0}{\hbar} \simeq 30.18$ nA. Note that the Andreev spectrum in the present model is a two-fold degenerate in the spin degrees of freedom because the CPR depends on the randomness of the on-site potentials. A typical result obtained in the presence of random on-site potentials is presented in Fig. 3(b) in the main text.

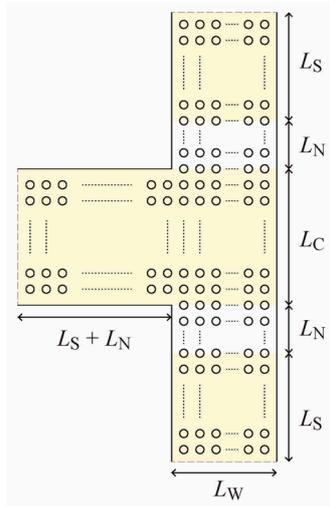

Fig. S5 Tight-binding model for coupled JJs described in the two-dimensional space. $L_W$, $L_C$, $L_N$, and $L_S$ are the widths of the upper and lower superconducting terminals, the width of the center superconducting terminal, the length of the normal regions, and the length of the upper and lower superconducting regions, respectively, with the lattice constant a.

**Supplementary Note 6: Potential tuning by the gate voltages**

The two-dimensional CPR $I_{sc1}(\phi_1, \phi_2)$ of JJ1 is strongly affected by the coherent coupling between the ABSs in JJ1 and JJ2. In the main text and Fig. 4, we discuss the gate voltage dependence of the two-dimensional CPR of JJ1. Negative gate voltages on both JJ1 and JJ2 push out the ABSs energetically, hence the coupling is reduced effectively. Here, we confirm it in the numerical calculation in Fig. S6. When the potential heights in the two normal regions in JJ1 and JJ2 are tuned in $1.5t \geq U_{g1}, U_{g2} \geq 0.5t$, the CPR $I_{sc1}(\phi_1, \phi_2)$ indicates a large modulation. At $U_{g1} = U_{g2} = 0.5t$, we obtain a large AJE in the CPR of JJ1. $U_{g1}$ acts as a barrier and reduces the Josephson current. We note that the potential heigh $U_{g1}$ ($U_{g2}$) increases when the gate voltage $V_{g1}$ ($V_{g2}$) is applied in negative. An extremly large $U_{g1}$ ($U_{g2}$) correspond to the situation that JJ1 (JJ2) are pinched off by the gate voltage. $U_{g1}$ becomes larger, some of the ABSs in JJ1 are energetically pushed out, but all the ABSs can couple with the ABSs in JJ2. Therefore, the AJE can be found even at $U_{g1} = 1.5t$ and $U_{g2} = 0.5t$. On the other hand, when the potential in JJ2 is higher, the smaller number of the JJ2 ABSs invokes that some of the JJ1 ABSs cannot couple with the JJ2 ABSs. Hence, the weaker AJE is obtained at larger $U_{g2}$. A whole feature of $I_{sc1}(\phi_1, \phi_2)$ in Fig. S6 qualitatively agrees with the experimental observation in Fig. 4. Therefore, our understanding of the AJE based on the coherent coupling between the ABSs is reasonable.

Here, we note validity of the parameter ranges in the numerical calculation. For the on-site potentials, we consider large changes in space such as $U_{g1}, U_{g2} \gtrsim t$ even though the superconducting pair potential is $\Delta_0 = 0.02t$. Such on-site potentials work as delta-function-like potential. For qualitative discussion, the delta-function potential explains the physics enough, e.g., in the Blonder-Tinkham-Klapwijk theory. Therefore, in the present discussion, we conclude that our numerical model has enough validity.

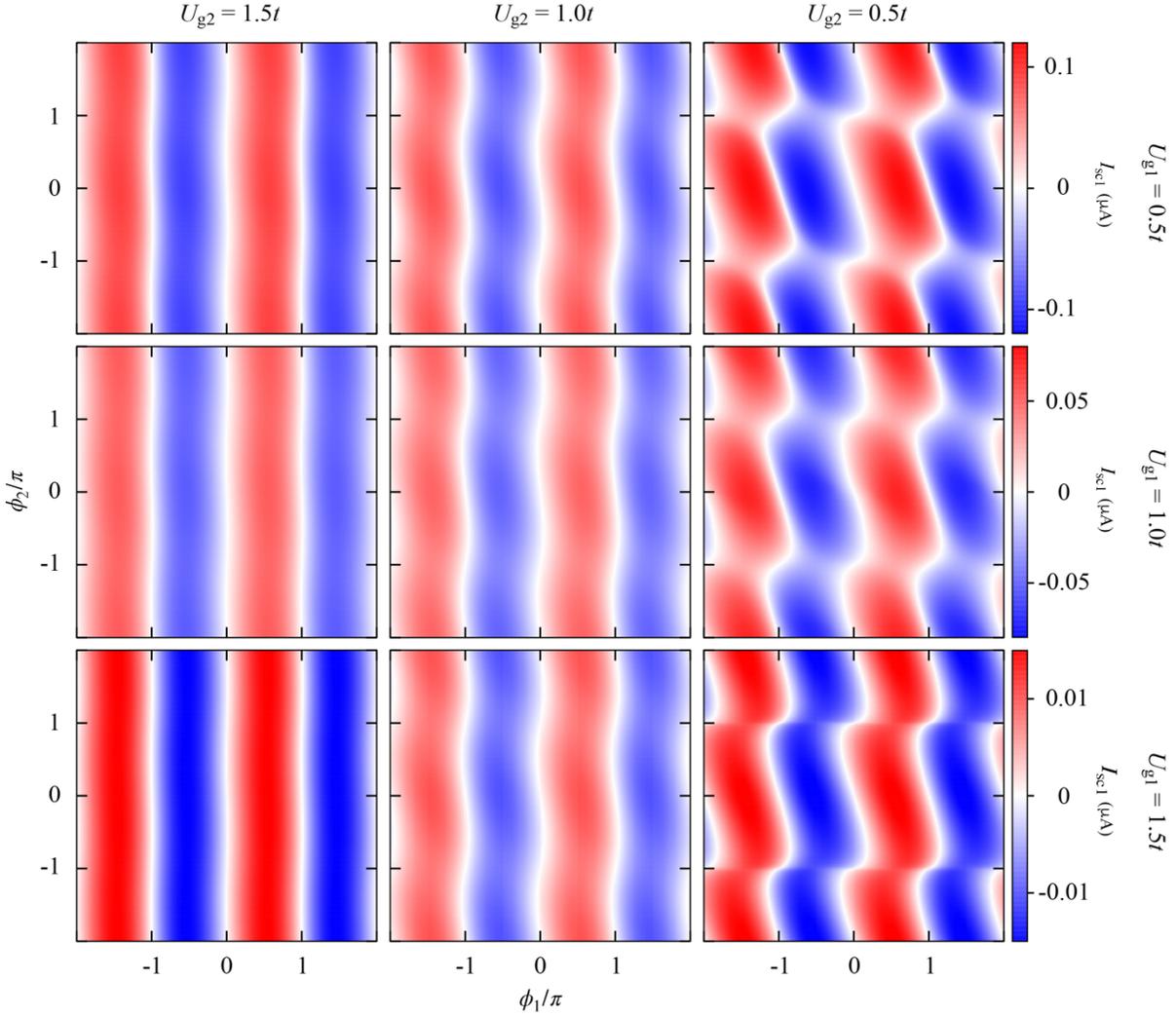

Fig. S6 : Two-dimensional CPR of JJ1, $I_{sc1}(\phi_1, \phi_2)$ when the potential heights in the normal regions of JJ1 and JJ2 are tuned. For lower potential in JJ2, the coherent coupling of the ABSs is stronger, hence the AJE is obtained even for large $U_{g1}$. For higher potential in JJ2, on the other hand, the AJE is not obtained even for high transparent JJ1 owing to a reduced coupling. To reduce a calculation cost, we set $N_S = 25$, which do not affect in the results.